\documentclass[conference,9pt]{IEEEtran}
\IEEEoverridecommandlockouts
\usepackage[T1]{fontenc}
\usepackage{amsmath,amssymb,amsthm,mathtools}
\usepackage{booktabs,array}
\usepackage{graphicx}
\usepackage{algorithm}
\usepackage{algpseudocode}
\algrenewcommand\algorithmicrequire{\textbf{Input:}}
\algrenewcommand\algorithmicensure{\textbf{Output:}}
\usepackage{microtype}
\usepackage{cite}
\usepackage[hidelinks]{hyperref}
\hypersetup{pdfauthor={Yunqi Huang, Wei Zi, Jianhao He, Zekun Ye, Shenggen Zheng},pdftitle={Ancilla-Free Exact Dicke-State Synthesis with Optimal Size and Depth}}
\newtheorem{theorem}{Theorem}
\newtheorem{lemma}{Lemma}
\newtheorem{proposition}{Proposition}

\theoremstyle{definition}
\newtheorem{problem}{Problem}
\theoremstyle{plain}
\newcommand{\ket}[1]{|#1\rangle}
\newcommand{\bra}[1]{\langle#1|}
\newcommand{\D}[2]{\ket{D_{#1}^{#2}}}
\newcommand{\norm}[1]{\lVert#1\rVert}
\newcommand{\Sn}{\mathcal S_n}
\newcommand{\sgn}{\operatorname{sgn}}
\title{Ancilla-Free Exact Dicke-State Synthesis\\with Optimal Size and Depth}
\author{%
\IEEEauthorblockN{Yunqi Huang\textsuperscript{1,*}, Wei Zi\textsuperscript{1,*},
Jianhao He\textsuperscript{2}, Zekun Ye\textsuperscript{3,$\dagger$},
and Shenggen Zheng\textsuperscript{1,$\dagger$}}
\IEEEauthorblockA{%
\textsuperscript{1}Quantum Science Center of Guangdong-Hong Kong-Macao Greater Bay Area, Shenzhen, China\\
\textsuperscript{2}Hetao Institute of Mathematics and Interdisciplinary Sciences, Shenzhen, China\\
\textsuperscript{3}Fuzhou University, Fuzhou, China}
\thanks{\textsuperscript{*}Yunqi Huang and Wei Zi contributed equally to this work.}
\thanks{\textsuperscript{$\dagger$}Corresponding authors: Zekun Ye (\href{mailto:yezekun@fzu.edu.cn}{\nolinkurl{yezekun@fzu.edu.cn}}) and
Shenggen Zheng (\href{mailto:zhengshenggen@quantumsc.cn}{\nolinkurl{zhengshenggen@quantumsc.cn}}).}}
\begin{document}
\bstctlcite{IEEEexample:BSTcontrol}
\maketitle
\begin{abstract}
We construct exact circuits for $n$-qubit Dicke states of any Hamming
weight $k$ using $O(n)$ single-qubit gates and CNOTs in $O(\log n)$
depth without ancillary qubits. The gate count and depth are asymptotically
optimal for every $0<k<n$. We compress symmetric blocks to free
output qubits for weight-dependent phases. A sign correction followed
by a collective rotation produces a state whose overlap with the target
is at least a positive constant, so exact amplitude amplification
takes $O(1)$ steps. For balanced Dicke states ($k=\lfloor n/2\rfloor$),
the logarithmic depth disproves the $\Omega(k)$ depth lower bound conjectured by Yuan
and Zhang (2025).
\end{abstract}

\section{Introduction}
The $n$-qubit Dicke state of weight $k$ is the uniform superposition
of all computational basis states with exactly $k$ ones,
\begin{equation*}
\D{k}{n}\coloneqq\binom nk^{-1/2}\sum_{|x|=k}\ket{x},
\qquad 0\le k\le n,
\end{equation*}
where $|x|=\sum_{i=1}^n x_i$ is the Hamming weight of
$x=(x_1,\ldots,x_n)\in\{0,1\}^n$.
These states form an orthonormal basis of the symmetric subspace~\cite{CGS}.
In the decoded quantum interferometry algorithm, $\D{k}{n}$ represents
a uniform superposition over all sets of $k$ error locations in a
codeword~\cite{DQI}. In quantum metrology,
the balanced state $\D{n/2}{n}$ serves as a probe for estimating small
collective spin rotations when $n$ is even~\cite{WangTerhal2021}.

For $0<k<n$, preparing $\D{k}{n}$ requires $\Omega(n)$ gates~\cite{WangMolmer2002,Huang2023}
and $\Omega(\log n)$ depth~\cite{YZ}.
When $r\coloneqq\min\{k,n-k\}$ is constant, the circuits of
B\"artschi and Eidenbenz~\cite{BE22} and Yuan and Zhang~\cite{YZ}
attain both bounds without ancillary qubits. For balanced Dicke states,
however, their bounds give $O(n^2)$ gates and $O(n)$ depth (Table~\ref{tab:prior}).
Liu, Childs, and Gottesman achieve polylogarithmic depth for arbitrary
weights by coherent symmetrization, using $O(n\log n)$ ancillary
qubits~\cite{LCG}. We ask whether every Dicke state admits a linear-size preparation
circuit that attains optimal depth using only the output qubits.

\begin{problem}[Exact Dicke-state preparation]\label{prob:dicke}
Can every $n$-qubit Dicke state be prepared exactly using $O(n)$ single-qubit gates and CNOTs,
with $O(\log n)$ depth and without ancillary qubits?
\end{problem}

\subsection{Our Results}
Our circuits meet the size and depth bounds in Problem~\ref{prob:dicke} using only the output qubits.

\begin{theorem}\label{thm:main}
For every $n\ge2$ and $1\le k\le n-1$, there is an ancilla-free
circuit that prepares $\D{k}{n}$ exactly using $O(n)$ arbitrary
single-qubit gates and CNOTs, with $O(\log n)$ depth.
\end{theorem}

In particular, for $k=\lfloor n/2\rfloor$, our logarithmic-depth
construction rules out the $\Omega(k)$ depth lower bound conjectured
by Yuan and Zhang~\cite{YZ} for Dicke-state preparation.
We call the Hamming weights $1\le k\le n-1$ \emph{nontrivial}.
For $k=0$, the input is already the target state. For $k=n$, we
prepare the target by applying a Pauli $X$ gate to each qubit in parallel.

For every nontrivial Dicke state, the reduced state of any pair of
qubits is entangled~\cite{WangMolmer2002}. Huang et al.~\cite{Huang2023}
call states with this property \emph{W-like} and show in their
Supplemental Material that preparing any such $n$-qubit state
requires $\Omega(n)$ CNOT gates. Yuan and Zhang~\cite{YZ} show that preparing any nontrivial
$n$-qubit Dicke state requires depth $\Omega(\log n)$. Thus our gate count and depth are both
asymptotically optimal.

\subsection{Related Work}
B\"artschi and Eidenbenz give exact Dicke-state preparation using $O(nr)$ gates and $O(n)$ depth without ancillary qubits~\cite{BE}. Their construction also yields coherent compression of symmetric states. Bellante et al.~\cite{RLS} develop TreeRLSP, a tree-tensor-network method for preparing regular language states, including Dicke states.

Liu, Childs, and Gottesman give deterministic coherent
symmetrization~\cite{LCG}. Zi, Yuan, and Nie~\cite{ZYN} start from a biased product state and
use an exact Hamming-weight counter to implement the reflections
for $O(r^{1/4})$ amplitude amplification steps. Our sign correction
and inverse rotation reduce the number of steps to $O(1)$.
Khattar et al.~\cite{DQI} use a reversible divide-and-conquer algorithm
to map each index to the corresponding set of occupied positions.
Their circuit uses $\widetilde O(n)$ gates for a sparse encoding of Dicke states.

Joshi and Vasconcelos give exact constant-depth Dicke-state preparation with unbounded fanout~\cite{JV}. Gretta, Gupta, and Joshi obtain constant-depth preparation of polylogarithmic-weight Dicke states using unbounded-control Toffoli gates, and of arbitrary symmetric states with additional fanout~\cite{GGJ}. Their Dicke-state construction combines constant initial overlap and uniform amplitudes in the target-weight subspace with exact
amplitude amplification.

Yu et al. use collective measurements and feedback~\cite{Yu}.
For even $n$, Yang, Wei, and Yin prepare $\D{n/2}{n}$ with an
adaptive circuit of depth $O(\log n)$ or a unitary circuit of depth
$O(\log^2 n)$~\cite{YWY}. Both use $O(n)$ additional qubits and
succeed with constant probability. They ask whether a unitary circuit
can achieve $O(\log n)$ depth with $O(n)$ ancillary qubits.
Theorem~\ref{thm:main} achieves this depth using only the output qubits.

\begin{table*}[t]
\centering
\caption{\normalfont
Selected upper bounds for deterministic exact preparation of a Dicke state of weight $k$ by unitary circuits.
Here $1\le k\le n-1$ and $r=\min\{k,n-k\}$.}
\label{tab:prior}
\begin{tabular}{lccc}
\toprule
Construction & Elementary gates & Depth & External ancillas\\
\midrule
B\"artschi--Eidenbenz~\cite{BE22}
& $O(nr)$
& $O(r\log(n/r))$
& $0$\\
Yuan--Zhang~\cite{YZ}
& $O(nr)$\textsuperscript{\ensuremath{\ddagger}}
& $O(r+\log(r+1)\log(n/r))$
& $0$\\
TreeRLSP~\cite{RLS}
& $O(n(r+1)^6)$
& $O((r+1)^6\log(n+1))$
& $0$\\
Zi--Yuan--Nie~\cite{ZYN}
& $O(nr^{1/4}\log(n+1))$
& $O(r^{1/4}\log(n+1))$
& $\lceil\log(n+1)\rceil$\\
Liu--Childs--Gottesman~\cite[Result~2]{LCG}
& $O(n\log^4(n+1)\log\log(n+1))$\textsuperscript{\ensuremath{\ddagger}}
& $O(\log^3(n+1)\log\log(n+1))$
& $O(n\log(n+1))$\\
\textbf{This work}
& $\boldsymbol{O(n)}$
& $\boldsymbol{O(\log(n+1))}$
& $\boldsymbol{0}$\\
\bottomrule
\end{tabular}
\par\smallskip
\begin{minipage}{\textwidth}
\footnotesize
\raggedright
For the entries marked \(\ddagger\), we bound the gate count by multiplying width and depth for the full circuit in~\cite{LCG} and for each subcircuit separately in~\cite{YZ}.
By~\cite[Lemma~12]{YZ}, we can limit each division subcircuit to
$O(r^2/\log(r+1))$ temporary qubits while preserving the depth bound.
This gives $O(r^2)$ gates per division subcircuit and $O(nr)$ gates
in total after summing over the $O(n/r)$ division and leaf subcircuits.
\end{minipage}
\end{table*}

\subsection{Construction Overview}
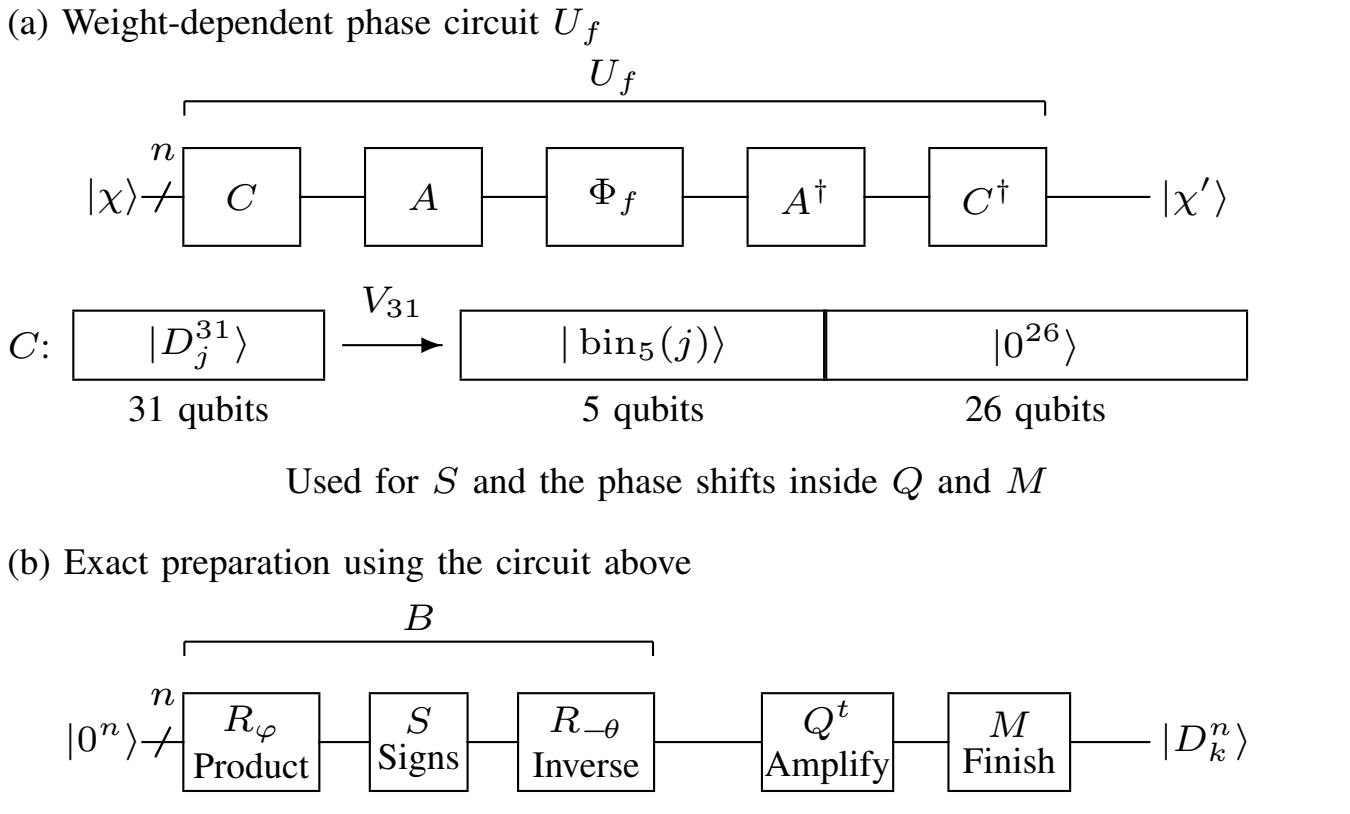
\begin{figure}[t]
\centering
\begingroup
\setlength{\unitlength}{0.9pt}
\scriptsize
\begin{picture}(276,177)
  \put(0,170){\makebox(276,7)[l]{(a) Weight-dependent phase circuit $U_f$}}
  \put(37,158){\line(1,0){180}}
  \put(37,155){\line(0,1){3}}
  \put(217,155){\line(0,1){3}}
  \put(37,159){\makebox(180,8){$U_f$}}
  \put(0,132){\makebox(28,12)[r]{$|\chi\rangle$}}
  \put(28,138){\line(1,0){9}}
  \qbezier(31,135)(32.5,138)(34,141)
  \put(28,144){\makebox(9,7){$n$}}
  \put(37,128){\framebox(24,20){$C$}}
  \put(61,138){\line(1,0){14}}
  \put(75,128){\framebox(24,20){$A$}}
  \put(99,138){\line(1,0){14}}
  \put(113,128){\framebox(28,20){$\Phi_f$}}
  \put(141,138){\line(1,0){14}}
  \put(155,128){\framebox(24,20){$A^\dagger$}}
  \put(179,138){\line(1,0){14}}
  \put(193,128){\framebox(24,20){$C^\dagger$}}
  \put(217,138){\line(1,0){22}}
  \put(241,132){\makebox(35,12)[l]{$|\chi'\rangle$}}
  \put(0,103){\makebox(12,8)[l]{$C$:}}
  \put(14,100){\framebox(52,14){$|D_j^{31}\rangle$}}
  \put(70,107){\vector(1,0){21}}
  \put(68,112){\makebox(25,8){$V_{31}$}}
  \put(95,100){\framebox(76,14){$|\operatorname{bin}_5(j)\rangle$}}
  \put(171,100){\framebox(88,14){$|0^{26}\rangle$}}
  \put(14,89){\makebox(52,8){31 qubits}}
  \put(95,89){\makebox(76,8){5 qubits}}
  \put(171,89){\makebox(88,8){26 qubits}}

  \put(0,74){\makebox(276,8){Used for $S$ and the phase shifts inside $Q$ and $M$}}
  \put(0,57){\makebox(276,8)[l]{(b) Exact preparation using the circuit above}}
  \put(37,45){\line(1,0){98}}
  \put(37,42){\line(0,1){3}}
  \put(135,42){\line(0,1){3}}
  \put(37,46){\makebox(98,8){$B$}}
  \put(0,18){\makebox(28,12)[r]{$|0^n\rangle$}}
  \put(28,24){\line(1,0){9}}
  \qbezier(31,21)(32.5,24)(34,27)
  \put(28,30){\makebox(9,7){$n$}}
  \put(37,14){\framebox(28,20){\shortstack{$R_{\varphi}$\\[-1pt]Product}}}
  \put(65,24){\line(1,0){11}}
  \put(76,14){\framebox(20,20){\shortstack{$S$\\[-1pt]Signs}}}
  \put(96,24){\line(1,0){11}}
  \put(107,14){\framebox(28,20){\shortstack{$R_{-\!\theta}$\\[-1pt]Inverse}}}
  \put(135,24){\line(1,0){23}}
  \put(158,14){\framebox(27,20){\shortstack{$Q^t$\\[-1pt]Amplify}}}
  \put(185,24){\line(1,0){12}}
  \put(197,14){\framebox(25,20){\shortstack{$M$\\[-1pt]Finish}}}
  \put(222,24){\line(1,0){17}}
  \put(241,18){\makebox(35,12)[l]{$|D_k^n\rangle$}}
\end{picture}
\endgroup
\caption{Circuit overview.
(a) For an input $\ket\chi\in\Sn$, $C$ compresses each block as illustrated. The qubits in state $\ket0$
are used to compute the total weight with $A$ and apply its phase with
$\Phi_f$. $A^\dagger$ uncomputes the arithmetic and $C^\dagger$
decompresses the blocks.
(b) For $1\le k\le n/2$, $B$ prepares the initial state,
$Q$ amplifies its target component, and $M$ completes phase matching
and global-phase correction (Section~\ref{sec:aa}).
Here $R_a\coloneqq R_y(a)^{\otimes n}$ and
$\varphi\coloneqq2\arcsin\sqrt{k/n}$.}
\label{fig:circuit-overview}
\end{figure}

Our construction has three components: a constant-overlap initial state,
ancilla-free weight-dependent phase circuits, and exact amplitude
amplification. Figure~\ref{fig:circuit-overview} shows how they fit together.

\emph{A constant-overlap initial state.}
After bit complementation, assume $1\le k\le n/2$.
One layer of single-qubit rotations prepares the biased product state
\begin{equation*}
\ket P\coloneqq(\sqrt{1-k/n}\ket0+\sqrt{k/n}\ket1)^{\otimes n}
       =\sum_w a_w\D{w}{n}.
\end{equation*}
Its overlap with $\D{k}{n}$ is $a_k=\Theta(k^{-1/4})$.
We choose the signs of the Dicke coefficients of $\ket P$ to match
those of a rotated Dicke state. Set
$R_\theta\coloneqq R_y(\theta)^{\otimes n}$ and
$\beta_w\coloneqq\bra{D_w^n}R_\theta\D{k}{n}$.
Apply a weight-dependent sign correction
$S\D{w}{n}\coloneqq\sgn(\beta_w)\D{w}{n}$, taking $\sgn(0)\coloneqq1$,
and then apply $R_\theta^\dagger$. The resulting target overlap is
\begin{equation*}
\bra{D_k^n}R_\theta^\dagger S\ket P=\sum_w a_w|\beta_w|.
\end{equation*}
For sufficiently large $k$, choose $\theta=\Theta(n^{-1/2})$.
Let $\sigma\coloneqq\sqrt{k(1-k/n)}$ denote the standard deviation
of the Hamming weight in $\ket P$. On a weight interval of width
$\Theta(\sigma)$ centered at $k$, every $a_w$ is
$\Omega(\sigma^{-1/2})$. We approximate the collective rotation by a quantum walk and show that the sum of the magnitudes of its amplitudes over this interval is $\Omega(\sqrt{\sigma})$. Multiplying these bounds gives a constant
overlap for the walk after sign correction. The approximation error is small enough that the overlap for the
collective rotation remains bounded below by a positive constant. For bounded $k$, the product
state alone suffices (Lemma~\ref{lem:overlap}).

\emph{Weight-dependent phases without ancillas.}
We need the phase circuit only on the symmetric subspace $\Sn$.
For such inputs, each full block of 31 qubits lies in the span of its 32 Dicke states.
Encoding the block weight in five qubits leaves the other 26 in $\ket0$.
We use these qubits to compute the total weight and apply the corresponding
phase, then reverse the arithmetic and decompress the blocks.
Lemma~\ref{lem:phase} thus implements any prescribed weight-dependent
phase on $\Sn$ with $O(n)$ gates and $O(\log n)$ depth using only
the output register. This gives both the sign correction $S$ and the
phase shifts needed for amplification.

\emph{Exact amplitude amplification.}
Let $B$ prepare the initial state above. Following the constant-overlap
approach of~\cite{GGJ}, we amplify the target amplitude by alternating
phase shifts on the target and on the initial state. The latter shift
uses $B$ and $B^\dagger$. Since the initial overlap is at least a positive constant, $O(1)$ iterations suffice. A final phase-matching
step cancels any remaining component orthogonal to $\D{k}{n}$. Each step
uses $O(n)$ gates and $O(\log n)$ depth, proving Theorem~\ref{thm:main}.

\section{Preliminaries}\label{sec:prelim}
\emph{States and notation.}
The Dicke states form an orthonormal basis
of the symmetric subspace
\begin{equation*}
\Sn\coloneqq\operatorname{span}\{\D{w}{n}:0\le w\le n\}.
\end{equation*}
For $0\le j<2^m$, define
$\operatorname{bin}_m(j)\coloneqq(j_{m-1},\ldots,j_0)\in\{0,1\}^m$ by
$j=\sum_{\ell=0}^{m-1}j_\ell 2^\ell$.
We use $\ket{\operatorname{bin}_m(j)}$ to denote the corresponding computational basis state. The symbol $\oplus$ denotes bitwise addition modulo two. All logarithms are base two.

\emph{Circuit model.}
We use arbitrary single-qubit gates and CNOTs. Circuit size counts these elementary gates. Depth counts
layers of gates acting on disjoint sets of qubits.
A Dicke-state preparation circuit maps $\ket{0^n}$ to $\D{k}{n}$ exactly.

For state preparation, ancillas are qubits beyond the $n$ output qubits.

\section{Weight-dependent phases within the output register}\label{sec:phase}

We construct an ancilla-free circuit that applies a prescribed phase
$e^{if(w)}$ to each Dicke component $\D{w}{n}$ of a symmetric state.
This operation is used for sign correction and amplitude amplification
in the state-preparation circuit.

We compress each constant-size symmetric block into a binary
encoding of its weight~\cite{BE}. The resulting qubits in state $\ket{0}$ are used to compute the
total weight by reversible carry-save addition~\cite{ZYNS}.
After applying the required phase, we reverse the computation and
decompress the blocks.

\begin{lemma}\label{lem:phase}
For every function $f:\{0,\ldots,n\}\to\mathbb R$,
there is an $n$-qubit circuit $U_f$ with $O(n)$ gates and
$O(\log(n+1))$ depth satisfying
\begin{equation*}
U_f\D{w}{n}=e^{if(w)}\D{w}{n},\qquad 0\le w\le n.
\end{equation*}
\end{lemma}

\begin{proof}
For $n\le31$, exact synthesis~\cite{Barenco} gives a circuit with constant gate count and depth. For $n\ge32$, partition the $n$ qubits into $Q\coloneqq\lceil n/31\rceil$ blocks of sizes $s_1,\ldots,s_Q$, where $s_i\coloneqq31$ for $i<Q$ and $s_Q\coloneqq n-31(Q-1)$. Set $b_i\coloneqq\lceil\log(s_i+1)\rceil\le5$.
Five qubits suffice to encode each block weight, leaving enough space
for the addition and phase circuits below.
For each block size $s$, choose a unitary $V_s$ satisfying
\begin{equation}\label{eq:compress}
V_s\D{j}{s}=\ket{\operatorname{bin}_{\lceil\log(s+1)\rceil}(j)} \ket{0^{s-\lceil\log(s+1)\rceil}}
\end{equation}
for every block weight $j\in\{0,\ldots,s\}$.
A single $V_s$ acts coherently on all these weights.
Since $s\le31$, each $V_s$ has an exact implementation with constant gate count and depth without ancillary qubits~\cite{Barenco,BE}. Applying these circuits to all blocks in parallel requires
$O(n)$ gates and constant depth.
The decomposition
\begin{equation*}
\D{w}{n}=\sum_{\sum_i j_i=w} \sqrt{\frac{\prod_i\binom{s_i}{j_i}}{\binom nw}} \bigotimes_i\D{j_i}{s_i}
\end{equation*}
allows each block to be compressed by $V_{s_i}$. Since block $i$ yields $s_i-b_i$ qubits in state $\ket{0}$ for every $j_i$, the compressed state contains $n-\sum_i b_i\ge n-5Q$ qubits in state $\ket{0}$. Summing the encoded block weights gives the total Hamming weight $w$.

\emph{Summation and storage.}
Treat the $Q$ binary block weights as rows and sum them using a tree
of reversible carry-save adders. Let \mbox{$m\coloneqq\lceil\log(n+1)\rceil$}.
Each full adder uses one qubit initialized to $\ket0$ to store the carry:
\begin{equation}\label{eq:adder}
(a,b,c,0)\longmapsto(a,b,a\oplus b\oplus c,ab\oplus ac\oplus bc).
\end{equation}
The identity $ab\oplus ac\oplus bc=a\oplus(a\oplus b)(a\oplus c)$ gives a constant-size implementation. First XOR $a$ into $b,c$ and compute their product into the carry qubit. Then XOR $a$ into that qubit, restore $b$, and XOR $b$ into the sum qubit.

The sum and carry satisfy $a+b+c=(a\oplus b\oplus c)+2(ab\oplus ac\oplus bc)$. Toffoli gates admit constant-size ancilla-free decompositions~\cite{Barenco}.

Missing bits are treated as zeros. A column with two input bits
uses one Toffoli gate to compute the carry and one CNOT to replace
the second bit by their XOR. Within each round, adders for different triples or columns act on disjoint qubits and run in parallel. We retain the input and carry bits needed for uncomputation.

For the $Q$ rows of at most five bits, let $M_t$ be the number of rows remaining to be summed after $t$ carry-save rounds with $M_0=Q$. Each round replaces $g_t\coloneqq\lfloor M_t/3\rfloor$ triples by pairs, so $M_{t+1}=M_t-g_t$. Let $T$ be the number of rounds needed to reduce the rows to two. Then
\begin{equation*}
M_{t+1}-2\le\tfrac23(M_t-2),\qquad
\sum_{t=0}^{T-1}g_t=Q-2.
\end{equation*}
Each round reduces the row count by $g_t$, so
$\sum_{t=u}^{T-1}g_t=M_u-2$. Summing over $u$ gives
\[
 \sum_{t=0}^{T-1}t g_t
 =\sum_{u=1}^{T-1}(M_u-2)
 \le (Q-2)\sum_{u=1}^{\infty}(2/3)^u=2(Q-2).
\]
Round $t$ has at most $5+t$ columns. The total number of carry qubits used across all rounds is at most
\begin{equation}\label{eq:csa}
\sum_t(5+t)g_t\le 7(Q-2).
\end{equation}
This gives $O(Q)$ gates and $O(\log Q)$ depth.

Add the two remaining numbers $A,B$ into an $m$-qubit register $W$ initialized to $\ket{0^m}$ using the carries
\begin{equation*}
c_0\coloneqq0,\qquad
c_{j+1}\coloneqq A_jB_j\oplus A_jc_j\oplus B_jc_j
\quad(0\le j<m-1)
\end{equation*}
and sum bits $W_j\coloneqq A_j\oplus B_j\oplus c_j$. Uncompute the carries while preserving $A,B$. This uses $m-1$ carry qubits and $O(m)$ gates and depth. Since $n+1\le32Q$, we have $m\le Q+4$. Together with the $5Q$ qubits initially used to encode the block weights and at most $7(Q-2)$ carry qubits retained from the carry-save additions, the total space is at most
\begin{equation}\label{eq:sumspace}
5Q+7(Q-2)+(2m-1)\le14Q-7\le n.
\end{equation}

\emph{Phase application.}
The total weight $w$ stored in $W$ determines the phase $e^{if(w)}$. The phase circuit reuses the qubits reset to $\ket{0}$ by uncomputing the temporary carry bits of the final ripple-carry adder. Split $W$ into three high-order bits and $\ell\coloneqq m-3$ low-order bits, and set $L\coloneqq2^\ell$. The encoded block weights and the carries retained from the carry-save additions together use at most $12Q-14$ qubits.

For each low-order bit $j$, a balanced CNOT tree creates $2^j-1$ additional computational-basis copies. The trees run in parallel and collectively use $L-1-\ell$ additional qubits in state $\ket{0}$ and $O(L)$ gates with depth $O(\ell+1)$. The copies are reused across all high-order cases and uncomputed after the last case.

For each high-order value, compute a flag indicating whether the three
high-order bits equal that value. This uses at most three available
qubits in state $\ket{0}$ for the root and intermediate flags.
Use the equality flag at the root to construct a binary prefix tree
for the low-order bits.
At a node, let $p$ be the flag and $y$ a copy of the corresponding
low-order bit of $W$. Let $q$ be an available qubit in state $\ket{0}$.
Apply a Toffoli from $p,y$ to $q$ followed by a CNOT from $q$ to $p$.
This replaces $(p,0)$ by $(p\wedge\neg y,p\wedge y)$.
The parent qubit $p$ is reused for the branch with $y=0$.
Thus a node is marked exactly when the high bits match and the
low bits read so far agree with its path. The $2^j$ nodes at level
$j$ use distinct copies of bit $j$, so they can be processed in parallel.
The tree has depth $O(\ell+1)$ and uses $L-1$ additional flag qubits.
Apply $\operatorname{diag}(1,e^{if(w)})$ at leaf $w$, taking the
phase angle to be zero for $w>n$. On each computational basis
input, only the leaf matching the full weight is marked across all
high-order cases.
Reverse the tree and the computation of the high-order equality flag
before the next high-order case.

The phase stage uses at most $3+(L-1)+(L-1-\ell)=2L+1-\ell$ temporary qubits
within the output register. Including $W$, the phase stage therefore occupies at most
\begin{equation}\label{eq:peak}
12Q-14+m+2L+1-\ell
=12Q+2L-10\le n.
\end{equation}
Indeed, $2L=2^{m-2}\le n/2$ and
$n\ge31Q-30\ge24Q-20$ for $Q\ge2$.
The maximum number of qubits used during addition is bounded separately by~\eqref{eq:sumspace}.
The same temporary qubits are reused in both stages.

Uncompute the arithmetic, then invert the block compressions.
The result is $e^{if(w)}\D{w}{n}$.
The high-order register has constant size, so the phase stage uses
$O(L)$ gates and depth $O(\ell+1)$. Together with compression and
arithmetic, this proves the stated bounds.
\end{proof}

To apply a phase to a single specified weight for $n\ge32$, test whether $W$ equals that weight. Apply one phase gate to the test flag and then reverse the test. The $m-1$ temporary carry qubits reset to $\ket{0}$ suffice, and the phase stage uses
$O(m)$ gates and depth. For weight zero, test that all encoded
block weights are zero using a balanced AND tree, apply the phase,
and reverse the test and compression. This avoids weight summation.
If $B_0$ is the number of bits encoding the block weights,
this requires $B_0-1$ available qubits in state $\ket{0}$, with
$2B_0-1\le10Q-1\le n$ for $n\ge32$.

\section{An initial state with constant overlap}\label{sec:seed}

Starting from a product state, we construct an initial state whose overlap with $\D{k}{n}$ is bounded below by a positive constant independent of $n$ and $k$. This bound allows us to prepare $\D{k}{n}$ exactly using $O(1)$ amplitude amplification steps.

Complementing all bits reduces nontrivial weights to $1\le k\le n/2$.
Set $p\coloneqq k/n$ and define the product state
\[
\ket P\coloneqq
(\sqrt{1-p}\ket0+\sqrt p\ket1)^{\otimes n}
=\sum_{w=0}^n a_w\D{w}{n},
\]
where $a_w\coloneqq\sqrt{\binom nw p^w(1-p)^{n-w}}$.
The probabilities $a_w^2$ form a binomial distribution with mean $k$
and standard deviation $\sigma\coloneqq\sqrt{k(1-k/n)}$.
Set $a_w\coloneqq0$ for integers $w<0$ or $w>n$, and let $a\coloneqq(a_w)_{w\in\mathbb Z}$.
Then $\norm a_2=1$. The standard deviation $\sigma$ of the weight
distribution satisfies $\sqrt{k/2}\le\sigma\le\sqrt k$. By Stirling's formula, the coefficient of $\D{k}{n}$
in $\ket P$ is $a_k=\Theta(\sigma^{-1/2})=\Theta(k^{-1/4})$.
Direct amplitude amplification therefore takes $\Theta(k^{1/4})$ iterations as $k$ grows.

\begin{figure}[t]
\centering
\includegraphics[width=\columnwidth]{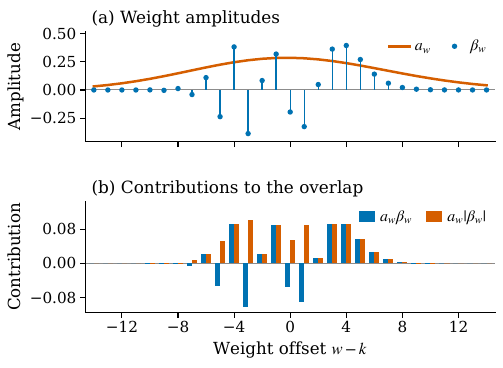}
\caption{For $n=128$ and $k=32$, the algorithm selects $d=1$,
giving $F_1\approx0.8384$. At this overlap, only the final phase-matching
step is needed. Starting from the product state ($F_0\approx0.2850$)
requires two standard amplification steps before phase matching. The plots show $|w-k|\le14$ at the selected angle.
(a) Product-state coefficients $a_w$ and rotated-target coefficients
$\beta_w$. (b) Contributions to the overlap after the inverse rotation,
$a_w\beta_w$ without sign correction and $a_w|\beta_w|$ with it.}
\label{fig:overlap}
\end{figure}

Let
\begin{equation*}
R_y(\alpha)\coloneqq
\begin{pmatrix}
\cos(\alpha/2)&-\sin(\alpha/2)\\
\sin(\alpha/2)&\cos(\alpha/2)
\end{pmatrix}
\end{equation*}
denote the single-qubit rotation about the $y$ axis, and recall that
$R_\theta=R_y(\theta)^{\otimes n}$. For $\theta\ge0$, let
\begin{equation*}
\beta_w\coloneqq\bra{D_w^n}R_\theta\D{k}{n},\qquad
S\D{w}{n}\coloneqq\sgn(\beta_w)\D{w}{n}.
\end{equation*}
We set the sign at zero to $+1$. Section~\ref{sec:classical} computes
$\sgn(\beta_w)$ exactly for the angles used below. Lemma~\ref{lem:phase} implements $S$ with $O(n)$ gates
and $O(\log(n+1))$ depth.
With $\varphi\coloneqq2\arcsin\sqrt p$, define
\begin{equation}\label{eq:B}
\begin{aligned}
B&\coloneqq R_{-\theta}S R_\varphi,\\
\ket\psi&\coloneqq B\ket{0^n}=R_{-\theta}S\ket P,\\
F&\coloneqq\bra{D_k^n}\psi\rangle=\sum_w a_w|\beta_w|.
\end{aligned}
\end{equation}
For $d\ge0$, let $F_d$ denote this overlap at
$\theta_d\coloneqq2\arctan(d/(2\sqrt n))$. The algorithm chooses the
largest overlap among $d\in\mathcal D\coloneqq\{0,1/4,1,2\}$ (Section~\ref{sec:classical}).
Figure~\ref{fig:overlap} illustrates the resulting sign correction.
The first two candidates suffice for the following bound.

\begin{lemma}\label{lem:overlap}
There is a constant $\eta>0$ such that
$\max\{F_0,F_{1/4}\}\ge\eta$ for all $1\le k\le n/2$.
\end{lemma}

For large $k$, we compare the rotated state at $d=1/4$ with a quantum
walk on the weight lattice. On an interval of width $\Theta(\sigma)$
around $k$, each $a_w$ is $\Omega(\sigma^{-1/2})$ and the absolute
walk amplitudes sum to $\Omega(\sqrt\sigma)$. Their weighted sum
therefore remains bounded below by a constant after accounting for the
approximation error. For bounded $k$, the product state suffices.

\subsection{Amplitudes of the Product State}
We first establish
\begin{equation}\label{eq:binomial}
a_k^2>\frac1{3\sigma},\qquad
a_{k+j}^2>\frac1{3\sigma}\quad(k\ge16,\ |j|\le\sigma/4).
\end{equation}
At $w=k$, Stirling's bounds give $a_k^2\ge e^{-1/6}/(\sqrt{2\pi}\sigma)>1/(3\sigma)$. 

For the second bound, set $w\coloneqq k+j$ and define
\begin{equation*}
g(w)\coloneqq w\ln(w/k)+(n-w)\ln((n-w)/(n-k)).
\end{equation*}
Stirling's bounds for $n!$, $w!$, and $(n-w)!$ give
\[
 a_w^2\ge
 \frac{\exp[-g(w)-1/(12w)-1/(12(n-w))]}{\sqrt{2\pi w(1-w/n)}}.
\]
Since $\sigma\le\sqrt k$ and $k\ge16$, both $w/k$ and $(n-w)/(n-k)$ lie in $[15/16,17/16]$. Thus $w,n-w\ge15$ and $w(1-w/n)\le(17/16)^2\sigma^2$. Throughout $|w-k|\le\sigma/4$, $g''(w)=1/w+1/(n-w)\le4/(3\sigma^2)$. Since $g(k)=g'(k)=0$, Taylor's theorem gives $g(w)\le2(w-k)^2/(3\sigma^2)\le1/24$. The Stirling correction for the denominator factorials is at most $1/(12w)+1/(12(n-w))\le1/90$.
Using $e^{-x}>1-x$ and $\sqrt{2\pi}<8/3$, we obtain
\begin{equation*}
a_w^2\ge\frac{e^{-19/360}}{(17/16)\sqrt{2\pi}\sigma} >\frac1{3\sigma}.
\end{equation*}

\subsection{Comparison with a Continuous-Time Quantum Walk}

\begin{figure}[t]
\centering
\begingroup
\setlength{\unitlength}{1pt}
\fontsize{7}{8}\selectfont
\begin{picture}(250,42)
  \put(0,28){\makebox(20,8)[l]{$H$}}
  \put(0,18){\makebox(20,7)[l]{$j=$}}
  \put(0,1){\makebox(20,8)[l]{$K$}}
  \put(23,28){\makebox(13,8){$\cdots$}}
  \put(223,28){\makebox(13,8){$\cdots$}}
  \put(23,1){\makebox(13,8){$\cdots$}}
  \put(223,1){\makebox(13,8){$\cdots$}}
  \put(36,32){\line(1,0){186}}
  \put(36,5){\line(1,0){186}}
  \put(42,32){\circle*{2.8}}
  \put(84,32){\circle*{2.8}}
  \put(126,32){\circle*{5}}
  \put(168,32){\circle*{2.8}}
  \put(210,32){\circle*{2.8}}
  \put(42,5){\circle*{2.8}}
  \put(84,5){\circle*{2.8}}
  \put(126,5){\circle*{5}}
  \put(168,5){\circle*{2.8}}
  \put(210,5){\circle*{2.8}}
  \put(32,18){\makebox(20,7){$-2$}}
  \put(74,18){\makebox(20,7){$-1$}}
  \put(116,18){\makebox(20,7){$0$}}
  \put(158,18){\makebox(20,7){$1$}}
  \put(200,18){\makebox(20,7){$2$}}
  \put(51,34){\makebox(24,8){$h_{-2}$}}
  \put(93,34){\makebox(24,8){$h_{-1}$}}
  \put(135,34){\makebox(24,8){$h_0$}}
  \put(177,34){\makebox(24,8){$h_1$}}
  \put(51,7){\makebox(24,8){$h$}}
  \put(93,7){\makebox(24,8){$h$}}
  \put(135,7){\makebox(24,8){$h$}}
  \put(177,7){\makebox(24,8){$h$}}
\end{picture}
\endgroup
\caption{Weight-basis comparison, with $j=w-k$.
The larger nodes mark the starting point $j=0$, corresponding to Hamming weight $w=k$.
Replacing $h_j$ by $h$ gives the walk used to compare
$|\beta_{k+j}|$ with $|J_j(z)|$.}
\label{fig:walk-comparison}
\end{figure}
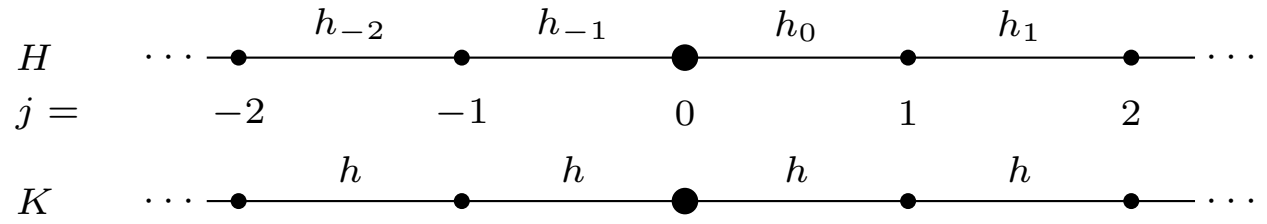

Near weight $k$, the rotation couples adjacent weights with nearly
constant strength. We approximate it by a continuous-time quantum walk
on $\mathbb Z$ with constant couplings.
For integer $j$, the Bessel function of the first kind has the integral
representation~\cite{DLMFBessel}
\begin{equation}\label{eq:bessel-integral}
J_j(z)\coloneqq\frac1\pi\int_0^\pi\cos(jt-z\sin t)\,dt.
\end{equation}
Set $h\coloneqq\sqrt{k(n-k)}/2$ and $z\coloneqq2h\theta$.
Extend $\beta_w$ by zero outside $0\le w\le n$.
For $\theta\ge0$, we claim
\begin{equation}\label{eq:duhamel}
\left(\sum_{j\in\mathbb Z}(|\beta_{k+j}|-|J_j(z)|)^2\right)^{1/2}
\le\frac{n\theta^2}{2\sqrt2}+\frac{3\sqrt n\,\theta}{2\sigma}.
\end{equation}
Identify $\ket j$ with $\D{k+j}{n}$ for $-k\le j\le n-k$.
Let $A$ be the generator of $R_\theta$
in this basis, extended by zero outside this range.
On $\ell^2(\mathbb Z)$,
define $D\ket j\coloneqq i^j\ket j$ and $T\ket j\coloneqq\ket{j+1}$ for $j\in\mathbb Z$.
The diagonal change of basis gives $D^\dagger A D=-iH$, where $H$
is a real symmetric tridiagonal operator with off-diagonal coefficients
\begin{equation*}
h_j\coloneqq\tfrac12\sqrt{(k+j+1)(n-k-j)}\quad(-k\le j<n-k),
\end{equation*}
and zero elsewhere. Compare $H$ with $K\coloneqq h(T+T^\dagger)$.
Figure~\ref{fig:walk-comparison} illustrates the two couplings.

For every integer $j$,
\begin{equation*}
|h_j-h|\le \lambda(|j|+1),\qquad
\lambda\coloneqq\frac{n}{2\sqrt{k(n-k)}}.
\end{equation*}
For $-k\le j<n-k$, subtract the expressions under the square roots and use $(k+j+1)(n-k-j)-k(n-k)=(n-2k-j)j+(n-k-j)$.
Here $|n-2k-j|\le n$ and $0<n-k-j\le n$.
Outside this range, $|j|+1\ge k$ and $\lambda k\ge h$.
Let $\phi(s)\coloneqq e^{-isK}\ket0$.
Under the Fourier transform, $K$ acts by multiplication by $2h\cos t$.
The Jacobi--Anger expansion~\cite{DLMFBessel} therefore gives
\begin{equation}\label{eq:walk-bessel}
\langle j|\phi(s)\rangle
=\frac1{2\pi}\int_{-\pi}^{\pi}e^{-i2hs\cos t}e^{ijt}\,dt
=(-i)^jJ_j(2hs).
\end{equation}
Applying Parseval's identity to $e^{-i2hs\cos t}$ and its derivative gives
$\norm{\phi(s)}_2=1$ and $\norm{M\phi(s)}_2=\sqrt2hs$,
where $(Mv)_j=jv_j$.
Shifting indices and applying the triangle inequality give
\begin{equation*}
\begin{aligned}
\norm{(H-K)\phi(s)}_2
&\le 2\lambda\norm{M\phi(s)}_2
   +3\lambda\norm{\phi(s)}_2\\
&=\lambda(2\sqrt2hs+3).
\end{aligned}
\end{equation*}
Since $H$ and $K$ are bounded and self-adjoint, Duhamel's formula gives
\begin{equation*}
\begin{aligned}
\norm{(e^{-i\theta H}-e^{-i\theta K})\ket0}_2
&\le\int_0^\theta\norm{(H-K)\phi(s)}_2\,ds\\
&\le\lambda(\sqrt2h\theta^2+3\theta).
\end{aligned}
\end{equation*}
Substituting $h$ and $\lambda$ gives~\eqref{eq:duhamel}, since taking
entrywise absolute values does not increase the $\ell^2$ distance. 

\subsection{Bounds on Bessel Functions}

Set $\tau\coloneqq11/12$ and
$K_\tau\coloneqq\sqrt{2/\pi}(1-\tau^2)^{-1/4}$.
Krasikov's bound~\cite{Krasikov2014} gives
\begin{equation*}
|J_j(z)|\le
\frac{\sqrt{2/\pi}}{(z^2-j^2+1/4)^{1/4}}
\qquad(j\ge1,\ z>\sqrt{j^2-1/4}).
\end{equation*}
Together with $|J_0(z)|\le\sqrt{2/(\pi z)}$
\cite{Krasikov2014} and
$|J_{-j}(z)|=|J_j(z)|$, this implies
\begin{equation}\label{eq:osc}
|J_j(z)|\le K_\tau/\sqrt z
\qquad(|j|\le\tau z).
\end{equation}
Parseval's identity applied to~\eqref{eq:walk-bessel} gives
$\sum_jJ_j(z)^2=1$ and $\sum_jj^2J_j(z)^2=z^2/2$. Hence
\begin{equation*}
\sum_{|j|\le\tau z}J_j(z)^2
\ge 1-\frac1{2\tau^2}=\frac{49}{121}.
\end{equation*}
The interval contains constant probability mass, while each
absolute amplitude is at most $K_\tau/\sqrt z$ by~\eqref{eq:osc}.
Hence
\begin{equation}\label{eq:bessel}
\sum_{|j|\le\tau z}|J_j(z)|
\ge \frac{49\sqrt z}{121K_\tau}.
\end{equation}

\begin{proof}[Proof of Lemma~\ref{lem:overlap}]
If $\sigma<48$, then $F_0=a_k>1/12$ by~\eqref{eq:binomial}.
Suppose $\sigma\ge48$, so $k\ge16$, and take $\theta=\theta_{1/4}$.
Since $x/2\le\arctan x\le x$ on $[0,1]$,
\[
  \tfrac18\le z/\sigma=\sqrt n\,\theta\le\tfrac14 .
\]
Thus $|j|\le\tau z$ implies $|j|\le\sigma/4$.
By~\eqref{eq:binomial}, \eqref{eq:bessel}, and $K_\tau<\sqrt2$,
\[
  \sum_j a_{k+j}|J_j(z)|
  \ge \frac{49}{121\sqrt3 K_\tau}\sqrt{\frac z\sigma}
  > \frac{1}{16\sqrt2}.
\]
By Cauchy--Schwarz and~\eqref{eq:duhamel}, this sum differs from
$F_{1/4}$ by at most $1/(32\sqrt2)+3/(8\sigma)$, since
$\norm a_2=1$. Hence
\[
 F_{1/4}>\frac{1}{32\sqrt2}-\frac{3}{8\sigma}>\frac{1}{128}.
\]
The two cases give the claim with $\eta=1/128$.
\end{proof}

\section{Exact amplitude amplification}\label{sec:aa}

\begin{proof}[Proof of Theorem~\ref{thm:main}]
By bit complementation, assume $1\le k\le n/2$. For $|u|=1$, let
$Z_w(u)$ multiply $\D{w}{n}$ by $u$ and fix all other Dicke basis
states. Lemma~\ref{lem:phase} implements this operation on $\Sn$.
For the angle selected in Section~\ref{sec:seed}, the state
$\ket\psi=B\ket{0^n}$ has overlap $F$ with $\D{k}{n}$.
If $F=1$, this state is already $\D{k}{n}$. Otherwise, put
$\alpha\coloneqq\arcsin F$ and write
\begin{equation*}
\ket\psi=F\D{k}{n}+\cos\alpha\ket{\perp},
\end{equation*}
where $\ket{\perp}$ is a unit vector orthogonal to $\D{k}{n}$.
The operator $BZ_0(u)B^\dagger$ applies phase $u$ to $\ket\psi$.
Thus the usual amplification step~\cite{BHMT} is
$Q\coloneqq-BZ_0(-1)B^\dagger Z_k(-1)$. After
$t\coloneqq\lfloor\pi/(4\alpha)-1/2\rfloor$ such steps,
\begin{equation*}
Q^t\ket\psi=\sin\beta\,\D{k}{n}+\cos\beta\ket{\perp},
\qquad \beta\coloneqq(2t+1)\alpha\le\pi/2.
\end{equation*}

If $\beta<\pi/2$, apply one phase-matching step~\cite{BHMT,Hoyer}.
After $Z_k(u)$, the overlap with $\ket\psi$ is
$\xi\coloneqq Fu\sin\beta+\cos\alpha\cos\beta$.
Applying $BZ_0(v)B^\dagger$ adds $(v-1)\xi\ket\psi$ to the state.
To cancel the coefficient of $\ket\perp$, we therefore need
\[
  v\xi=\xi-\frac{\cos\beta}{\cos\alpha}.
\]
Since $|v|=1$, the right-hand side must have modulus $|\xi|$. Squaring and substituting the expression for $\xi$ gives
\[
  2\cos\alpha\operatorname{Re}\xi=\cos\beta,
  \qquad \operatorname{Re}u=-\cot(2\alpha)\cot\beta .
\]
For $\alpha<\pi/4$, the choice of $t$ gives
$\beta>\pi/2-2\alpha$, so the required real part lies in $(-1,0)$.
For $\alpha\ge\pi/4$, we have $t=0$ and it equals
$1-1/(2F^2)\in[0,1/2)$. Choose $u$ on the upper unit semicircle
with this real part. Then $\operatorname{Re}\xi>0$, and the required
second phase is $v=-\overline{\xi}/\xi$.
The coefficient of $\D{k}{n}$ is then
$u\sin\beta-\tan\alpha\cos\beta$. If $\beta=\pi/2$, the coefficient
is already one after $Q^t$. In either case a single-qubit gate
proportional to the identity cancels the final global phase, including
the sign in $Q$ if omitted from its implementation.

Each complete weight-dependent phase circuit and each collective
rotation preserves $\Sn$. Lemma~\ref{lem:overlap} gives $F\ge\eta$,
so the number of amplification steps is $O(1)$.
The circuit $B$ consists of one phase circuit and two layers of
single-qubit rotations. Each amplification step uses $B$, $B^\dagger$
and two single-weight phase circuits. Lemma~\ref{lem:phase} therefore
gives $O(n)$ gates and $O(\log n)$ depth without ancillary qubits,
including any final bit flips.
\end{proof}

\section{Classical circuit generation}\label{sec:classical}

Algorithm~\ref{alg:generation} generates the circuit by selecting
the rotation with the largest overlap after sign correction. It compares
the four squared overlaps exactly and keeps the smallest $d$ in case of
a tie. Both~\eqref{eq:recurrence} and~\eqref{eq:Falg} use the target
weight $r=\min\{k,n-k\}$, with $\sgn(0)=1$.

\begin{algorithm}[tb]
\caption{Generate a circuit for $\D{k}{n}$}\label{alg:generation}
\small
\algtext*{EndIf}
\begin{algorithmic}[1]
\Require Integers $n\ge 2$ and $0\le k\le n$
\Ensure Layers of CNOTs and exact algebraic single-qubit gates
with their qubit indices
\State \algorithmicif\ $k=0$ \algorithmicthen\ \Return the empty circuit
\State \algorithmicif\ $k=n$ \algorithmicthen\ \Return one layer of $n$ $X$ gates
\State $r\gets\min\{k,n-k\}$, \quad $p\gets r/n$
\State $d_\star\gets 0$, \quad $F\gets\sqrt{\binom nr p^r(1-p)^{n-r}}$, \quad
$s\gets(+1)_{w=0}^n$
\For{$d$ in $(1/4,1,2)$}
  \State Compute $H_0,\ldots,H_n$ by~\eqref{eq:recurrence} with $q=4n/d^2$
  \State Compute $F_d$ by~\eqref{eq:Falg}
  \If{$F_d^2>F^2$}
    \State $(d_\star,F,s)\gets(d,F_d,(\sgn H_w)_{w=0}^n)$
  \EndIf
\EndFor
\State Generate the sign circuit $S$ for $s$ using Lemma~\ref{lem:phase}
\State $\varphi\gets 2\arcsin\sqrt p$, \quad
$\theta\gets 2\arctan(d_\star/(2\sqrt n))$
\State Form $B$ from $R_\varphi$, $S$, and $R_{-\theta}$, in that order
\State $C\gets B$
\If{$F<1$}
  \State Append to $C$ the amplification steps for $B,F,r$ (Section~\ref{sec:aa})
\EndIf
\State \algorithmicif\ $k>n/2$ \algorithmicthen\ append $X^{\otimes n}$ to $C$
\State \Return $C$
\end{algorithmic}
\end{algorithm}

\begin{proposition}\label{prop:classical}
Algorithm~\ref{alg:generation} generates the circuits of
Theorem~\ref{thm:main} in time polynomial in $n$.
\end{proposition}
\begin{proof}
An integer recurrence determines every required sign exactly. Each
overlap $F_d$ has algebraic degree at most four, and all $F_d^2$ lie in a
common real field of degree at most two. As in Section~\ref{sec:seed},
assume $1\le k\le n/2$ after bit complementation.

\emph{Exact signs.}
Fix $d>0$ in $\mathcal D$ and suppress its subscript for the moment.
Set $q\coloneqq 4n/d^2$, an integer for each of the three positive candidates.
For any output string of weight $w$, exactly
$\binom wj\binom{n-w}{k-j}$ weight-$k$ input strings share $j$
ones with it. Each has $k-j$ transitions $1\to0$, giving the
sign $(-1)^{k-j}$, and $n-w-k+2j$ unchanged bits.

Each unchanged
bit contributes $\sqrt q$ after the common factor
$(q+1)^{-n/2}$ is taken out of the product. Summing over $j$ and
normalizing the Dicke states gives
\begin{equation*}
\beta_w=\sqrt{\frac{\binom nw}{\binom nk}}
\frac{q^{(n-w-k)/2}}{(q+1)^{n/2}}H_w,
\end{equation*}
where
\begin{equation*}
H_w\coloneqq\sum_{j=\max\{0,k-(n-w)\}}^{\min\{w,k\}}
(-1)^{k-j}\binom wj\binom{n-w}{k-j}q^j.
\end{equation*}
Thus $\sgn(\beta_w)=\sgn(H_w)$, with $\sgn(0)=1$.
Starting from $H_0=(-1)^k\binom nk$, the Krawtchouk difference
relation~\cite{DLMF} yields
\begin{equation}\label{eq:recurrence}
\begin{aligned}
(n-w)H_{w+1}&=[n-w+qw-k(q+1)]H_w\\
&\hspace{1em}-qwH_{w-1}\quad(0\le w<n).
\end{aligned}
\end{equation}
The last term is omitted at $w=0$. Indeed,
$H_w=[t^k]P_w(t)$ for $P_w(t)\coloneqq(1+qt)^w(1-t)^{n-w}$,
and taking the coefficient of $t^k$ in
\begin{equation*}
(n-w)P_{w+1}+qwP_{w-1}
=(n-w+qw)P_w-(q+1)tP'_w
\end{equation*}
gives~\eqref{eq:recurrence}. At $d=0$ the rotation is the identity,
so $\beta_w=\mathbf1_{\{w=k\}}$ and every sign is $+1$.

\emph{Overlaps and angle selection.}
For $d>0$, write
\begin{equation*}
b\coloneqq\frac{p}{q(1-p)},\qquad
A\coloneqq\frac{(1-p)^n q^{n-k}}{\binom nk(q+1)^n}.
\end{equation*}
Separating the even and odd weights in~\eqref{eq:B} gives
\begin{equation}\label{eq:Falg}
\begin{aligned}
F_d&=\sqrt A(E_0+E_1\sqrt b),\\
E_j&\coloneqq\sum_{h=0}^{\lfloor(n-j)/2\rfloor}
 \binom n{2h+j}|H_{2h+j}|b^h\qquad(j=0,1).
\end{aligned}
\end{equation}
Here $A,b,E_0,E_1$ are rational. Because
$\sqrt b=d\sigma/[2(n-k)]$, every $F_d^2$ lies in $\mathbb Q(\sigma)$. For $d=0$,
$F_0^2=a_k^2=\binom nk p^k(1-p)^{n-k}$ is rational.
All $F_d$ are nonnegative, so comparing their squares in
$\mathbb Q(\sigma)$ determines the maximizing candidate exactly.

Vandermonde's identity gives $|H_w|\le q^k\binom nk\le q^k2^n$.
These integers and the numerators and denominators of $A,b,E_0,E_1$
have $O(n\log(n+1))$ bits. Generating the binomial coefficients
consecutively and evaluating the parity sums by Horner's rule takes
$O(n)$ arithmetic operations per candidate, including the recurrence.
Fast integer arithmetic therefore computes all sign tables and
overlaps in $\widetilde O(n^2)$ bit operations.

\emph{Amplification parameters.}
For the selected overlap $F<1$, compute $\cos\alpha=\sqrt{1-F^2}$.
If $F^2>1/2$, then $t=0$. Otherwise start with $(x,y)=(F,\cos\alpha)$
and repeatedly form
\begin{equation*}
\binom{x'}{y'}=
\begin{pmatrix}
1-2F^2 & 2F\cos\alpha\\
-2F\cos\alpha & 1-2F^2
\end{pmatrix}\binom{x}{y}.
\end{equation*}
This adds $2\alpha\le\pi/2$ to the current angle. Accept the
update when $y'\ge0$, and stop when $y'=0$ or before the first
update with $y'<0$. The number of accepted updates is $t$, and the
final pair is $(\sin\beta,\cos\beta)$. For $t=0$ this is the
initial pair. Since $F\ge\eta$, only $O(1)$ comparisons are needed.

The formulas in Section~\ref{sec:aa} now give the final phases by
exact algebraic operations. Adjoining $\cos\alpha$,
$\sqrt{1-(\operatorname{Re}u)^2}$ and $i$ to $\mathbb Q(F)$ gives
a field of degree at most 32. The constant number of operations
preserves the $O(n\log(n+1))$ bound on coefficient bit lengths,
so bounded-degree arithmetic and sign tests take polynomial time.

\emph{Elementary gates.}
For each block size $s\le31$, compose the inverse coherent
Dicke transform~\cite{BE} with a unary-to-binary conversion.
The resulting circuit acts only on the \(s\) qubits of the block, and the entries of each gate matrix are algebraic numbers.
For $n\le31$, exact synthesis of the phase circuit uses $O(1)$ algebraic
operations and root extractions.
These constructions generate the qubit indices, parallel layers and exact algebraic
descriptions of the gate matrices in polynomial time in $n$.
\end{proof}

\section{Conclusion}
We give ancilla-free circuits that exactly prepare any $n$-qubit Dicke
state using $O(n)$ gates in $O(\log n)$ depth. Both bounds are
asymptotically optimal for $0<k<n$. By implementing weight-dependent
phases within the output register, we prepare an initial state with
constant overlap with the target, so exact amplitude amplification
takes only $O(1)$ steps. Future work includes reducing the gate count by optimizing the phase circuits and the rotation angle,
and adapting the construction to finite gate sets and restricted
qubit connectivity.

\section*{Acknowledgments}
Yunqi Huang and Shenggen Zheng were supported by the Guangdong Provincial
Quantum Science Strategic Initiative (Grant No.~GDZX2503001) and the
Shenzhen Science and Technology Program (Grant No.~ZDYJ20251211120859002).
Zekun Ye was supported by start-up research funding from Fuzhou University
(Grant No.~XRC-26004). Wei Zi acknowledges support from the National Natural Science Foundation of China (Grant No. 62602563).

\section*{Content Generated by AI}
OpenAI ChatGPT (Codex) assisted with the Abstract and Sections I--VII:
literature comparison, proof development and checking, drafting and
revision, the figures, and resource accounting.
Its contributions include generated mathematical
arguments and code. The authors are responsible for all claims,
references, and results.

\clearpage
\begingroup
\let\footnotesize\small
\IEEEtriggeratref{12}
\bibliographystyle{IEEEtran}
\bibliography{ref}
\endgroup
\end{document}